\documentclass[twocolumn,showpacs,preprintnumbers,amsmath,amssymb,epsf]{revtex4-1}

\usepackage{graphicx,epsfig}
\usepackage{dcolumn}
\usepackage{bm}
\newcommand{\be}{\begin{equation}}
\newcommand{\ee}{\end{equation}}
\newcommand{\bse}{\begin{subequations}}
\newcommand{\ese}{\end{subequations}}
\newcommand{\bary}{\begin{eqnarray}}
\newcommand{\eary}{\end{eqnarray}}
\newcommand{\bwt}{\begin{widetext}}
\newcommand{\ewt}{\end{widetext}}

\begin{document}


\title{Superluminal Neutrinos in a Pseudoscalar Potential}
\author{Sarira Sahu$^{*}$, Bing Zhang$^{**}$}
\affiliation{
$^{*}$Instituto de Ciencias Nucleares, Universidad Nacional Aut\'onoma de M\'exico, 
Circuito Exterior, C.U., A. Postal 70-543, 04510 Mexico DF, Mexico\\
$^{**}$Department of Physics and Astronomy, University of Nevada, Las Vegas, NV 89154, USA
}

\email{Email: sarira@nucleares.unam.mx, zhang@physics.unlv.edu}
\begin{abstract}
The superluminal propagation of neutrinos observed 
by OPERA collaboration can be interpreted as neutrinos traveling
in a pseudoscalar potential which may be generated by a
medium. The OPERA differential arrival time data 
set a constraint on the form of the pseudoscalar potential. 

\end{abstract}

\maketitle

\section{Introduction}

The OPERA collaboration recently reported the evidence of superluminal
propagation of muon neutrinos \cite{OPERA11}. An earlier report
by the MINOS collaboration \cite{adamson07} reached the similar
conclusion with a lower statistical confidence.
If the result is not
due to an unknown instrumental effect, this fact inevitably points
towards new physics. Since the release of the OPERA data, many ideas
have been disposed to interpret the data, e.g. 
\cite{amelino-camelia11,tamburini11,giudice11,dvali11,gubser11,franklin11}.
The difficulties of various interpretations have been discussed
\cite{drago11,magueijo11,cohen11,ciuffoli11,bi11}. 
One of the difficulties is to account for the energy
dependence of neutrino speed anomaly $\Delta = (v-c)/c$. While 
OPERA reports 
\bwt
\begin{eqnarray}
\Delta = (2.48\pm 0.28({\rm stat.})\pm 0.30({\rm sys.})\times 10^{-5},
& \bar E_\nu = 17~{\rm GeV}~ \\
\Delta_1 = (2.18\pm 0.77({\rm stat.})\pm 0.30({\rm sys.})\times 10^{-5},
& ~\bar E_{\nu,1} = 13.9~{\rm GeV}~ \label{Delta1} ~\\
\Delta_2 = (2.76\pm 0.75({\rm stat.})\pm 0.30({\rm sys.})\times 10^{-5},
& ~\bar E_{\nu,2} = 42.9~{\rm GeV}~ \label{Delta2}
\end{eqnarray}
\ewt
for the anomaly in the entire sample, $E_\nu < 20$ GeV, and $E_\nu > 20$
GeV samples, respectively, SN 1987A neutrino arrival data poses a
stringent constraint
\begin{eqnarray}
|\Delta| < 2\times 10^{-9}, & ~\bar E_\nu = 10~{\rm MeV}~.
\end{eqnarray}
The MINOS data, with less significance, suggests
\begin{eqnarray}
\Delta = (5.1 \pm 2.9) \times 10^{-5}, & ~\bar E_\nu = 3~{\rm GeV}~.
\end{eqnarray}
This further sharpens the $\Delta$ contrast in the 10 MeV to 10
GeV range. 

Here we propose a superluminal neutrino theory without introducing
Lorentz Invariance Violation (LIV) or particles with imaginary masses
(tachyons). We envisage a pseudoscalar potential under the influence 
of which the neutrino propagates from CERN to OPERA detector.


\section{Superluminal Neutrinos in a Pseudoscalar Potential}

We introduce a pseudoscalar potential  $\phi > 0$,
which can be energy-dependent, but is constant in space (or can be
approximated as a constant for the spatial range in consideration).
The Dirac equation for a neutrino with mass $m_{\nu}$ in such a 
pseudoscalar potential can be written as
\be
\left [   {\bf \alpha . p} + \beta (m_{\nu}+ \gamma_5 \phi)  \right ]\psi =
E_{\nu} \psi,
\label{dirac}
\ee
where $\alpha$, $\beta$ and $\gamma_5$ are the Dirac matrices. 
The natural units with $c=1$ and $\hbar=1$ are adopted here.
The neutrino wave function $\psi$ can be expressed in terms of 
two-component spinors $u$ and $v$  as
\be
\psi= N \left (
   \begin{array}{c}
   u \\
   v
   \end{array}
   \right ), 
\ee
where $N$ is the normalization constant. These two spinors $u$
and $v$ satisfy the coupled equations
\be
   \begin{array}{c}
   \left ( \sigma . p - \phi  \right ) u- (E_{\nu}-m_{\nu}) v=0\\
   \left ( \sigma . p + \phi \right ) v - (E_{\nu}+m_{\nu}) u=0
   \end{array}.
\ee
Solving these two equations, we obtain the neutrino energy
\be
E_{\nu}^2={\bf p}^2+m^2_{\nu}-\phi^2.
\label{E-p}
\ee
For $m_{\nu} \ll \phi \ll E_{\nu}$, one can write
\be
|{\bf p}| \simeq E_{\nu}+\frac{\phi^2}{2 E_\nu^2}~.
\ee
The group velocity of the neutrino is then
\be
v_g=\frac{dE_\nu}{d|{\bf p}|}=
1+\frac{\phi^2}{2 E_{\nu}^2}-\frac{\phi \dot \phi}{E_{\nu}}~,
\ee
where  $\dot \phi={d\phi} / {dE_{\nu}}$, or
\begin{equation}
\Delta = v_g -1 = \frac{\phi^2}{2 E_{\nu}^2}-\frac{\phi \dot \phi}{E_{\nu}}~.
\end{equation}
The condition of superluminal neutrinos ($\Delta > 0$) can be written as
\be
\frac{\phi}{E_{\nu}}  > 2 \dot \phi.
\label{condition}
\ee

\section{Constraining the Form of $\phi$ with Data}

The condition (\ref{condition}) is satisfied for a wide range of 
$\phi$ forms. Below we attempt to construct an analytical form
of $\phi$ that gives superluminal motion and satisfies the 
observational constraints.

\subsection{Constant $\phi$}

An energy-independent potential ($\dot \phi = 0$)
naturally satisfies the condition (\ref{condition}), so that 
\begin{equation}
\Delta = v_g - 1 =\frac{|{\bf p}|}{E_{\nu}}-1 = \frac{\phi^2}{2 E_{\nu}^2} > 1~.
\end{equation}
Such a form is similar to many theories invoking LIV or tachyon 
particles e.g. \cite{drago11,bi11,franklin11}, which is ruled out
by the OPERA differential speed data. 
The ratio of velocity difference in this model is
$\Delta_1/\Delta_2 = (\bar E_{\nu,2}/\bar E_{\nu,1})^2$.
According to equations (\ref{Delta1})
and (\ref{Delta2}), the ratio between two average neutrino energies is
about $\bar E_{\nu,2}/\bar E_{\nu,1} \sim 3.1$. This corresponds to 
a predicted $\Delta_1/\Delta_2 \sim 9.5$, while the observed ratio
is $\Delta_1/\Delta_2 = 0.79^{+1.11}_{-0.50}$.
So a more complicated form of $\phi$ with a much shallower $E_\nu$ 
dependence on $\Delta$ is needed.

\subsection{Power law}

For a pseudoscalar potential of the form
\be
\phi(E_{\nu}) = A  E_{\nu}^{\alpha},
\ee
the superluminal condition (\ref{condition}) can be translated to
\be
-0.5 < \alpha < 0.5.
\label{PLcond}
\ee
The ratio of velocity difference in this model is $\Delta_1/\Delta_2
=(\bar E_{\nu,1}/\bar E_{\nu,2})^{2\alpha-2}$. Solving for $\alpha$
using the OPERA data, one gets
\be
0.72 < |\alpha| < 1.55
\ee
with the typical value $|\alpha| \sim 1.1$. This range is inconsistent
with the condition (\ref{PLcond}), and is in the subluminal regime.
Therefore the power law model cannot interpret the OPERA data.

\subsection{Power law exponential}

The pseudoscalar potential of the form
\be
\phi(E_{\nu}) = B {\left(\frac{E_{\nu}}{E_{0}}\right)}^{\alpha} 
e^{-E_\nu/E_{0}}
\ee
gives
\be
\Delta = \frac{\phi^2}{{E_\nu}^2} \left(\frac{1}{2}-\alpha+\frac{E_{\nu}}
{E_{0}}\right).
\ee
The superluminal condition is $\alpha < 1/2 + E_\nu/E_{0}$. However,
when $E_\nu \ll E_{0}$ and $E_\nu \gg E_{0}$, the velocity
difference has the energy-dependence in the form of 
$\Delta \propto E_{\nu}^{-2}$ and $\Delta \propto E_{\nu}^{-1}$,
respectively. Both dependences are too steep to account for the OPERA
data in equations (\ref{Delta1}) and (\ref{Delta2}). This form is
therefore not favored.

\subsection{Power law logarithmic}

We consider the potential of the form
\be
\phi(E_{\nu})= C {\left(\frac{E_{\nu}}{E_{0}}\right)}^{\alpha} 
\ln \left(\frac{E_\nu}{E_{0}}\right)~.
\label{PLlog}
\ee
The superluminal condition (\ref{condition}) can be translated to
\be
\frac{E_\nu}{E_{0}} > \exp \left({\frac{2}{1-2\alpha}}\right)~
\label{PLlogcond}
\ee
for $\alpha < 1/2$ (the branch $\alpha > 1/2$ is not favored since
the $E_\nu$-dependence of $\Delta$ is very steep).
The velocity difference takes the form
\be
\Delta = \frac{C^2}{E_\nu^2} \left(\frac{E_{\nu}}{E_{0}}\right)
^{2\alpha} \ln \left(\frac{E_{\nu}}{E_{0}}\right)
\left[\left(\frac{1}{2}-\alpha\right) \ln \left(\frac{E_\nu}
{E_{0}}\right) -1 \right]~.
\label{PLlogDelta}
\ee
For a not very high $E_{0}$, the superluminal
condition (\ref{PLlogcond}) is easily satisfied.
The steep $\propto E_{\nu}^{-2}$ dependence is
compensated by other factors in equation (\ref{PLlogDelta}) so
that a shallow $E_\nu$-dependence can be achieved in the 
superluminal regime. By properly adjusting $E_0$ and normalization
$C$, the OPERA data can be interpreted.


\section{Discussion}

We have shown that if neutrinos travel in a pseudoscalar potential,
superluminal propagation is possible under the condition given in
equation (\ref{condition}),
without violating special relativity. 
In order to interpret the OPERA data, a shallow
energy-dependence is required in the superluminal regime.
A potential form similar to equation (\ref{PLlog}) with 
$\alpha < 1/2$ can meet such a requirement.

The very small $|\Delta|$ derived for SN 1987A of 10 MeV neutrinos
is difficult to account for with such a potential. If one adjusts
$E_0$ to interpret OPERA data, the 10 MeV neutrinos would be in
the subluminal regime with a large $|\Delta|$ violating the data
constraint. If one instead adjusts 10 MeV to the transition
point from superluminal to subluminal,
i.e. $E_{\nu} \sim E_{\nu,c}=E_{0} \exp(\frac{2}{1-2\alpha})
\sim 10$ MeV, then the energy-dependence in the 10 GeV range
is too steep to satisfy the OPERA data. 
In order to incorporate the SN 1987A data, one needs to either
argue for a local effect of OPERA anomaly (e.g. the pseudoscalar
potential is related to the density, gravity or magnetic fields 
in the neutrino propagation path in the earth crust)
or admit that the potential (\ref{PLlog}) is not the correct 
form in the low energy regime.


If, however, the pseudoscalar potential (\ref{PLlog}) can be
extended to the high energy regime, it is interesting to note
that in the $\sim$ PeV energy range where the internal-shock-origin 
neutrinos from gamma-ray bursts (GRBs) are supposed to be generated
\cite{waxman97}, neutrinos are still superluminal but with
a much smaller $\Delta \sim 10^{-10}$. For typical 
high-luminosity GRBs at cosmological distances, 
the lead time would be several years.
For nearby low-luminosity GRBs \cite{gupta07,murase06}, the lead
time can be as short as several months. This can be in principle
tested by independent discoveries of a high-energy neutrino burst
detected by IceCube and a later GRB detected in the same direction.
This may be possible for a bright, nearby GRB event such as GRB 030329. 
The TeV neutrinos \cite{meszaros01,razzaque03}
have a much larger $\Delta \sim 10^{-8}$, and hence, a leading time
of decades to centuries, which is difficult to test observationally.

Finally, we'd like to comment on that the normalization parameter 
$C$ in (\ref{PLlog}) adjusted to fit the OPERA data naturally
leads to $\phi \gg m_\nu$, but $\phi \ll m_e$. So it is allowed
that the Dirac equation with pseudoscalar potential (\ref{dirac})
also applies to electrons and other spin 1/2 leptons, but those
particles cannot be superluminal since their masses dominate
the $\phi$ term in the energy-momentum equation (\ref{E-p}).


\begin{thebibliography}{55}
\expandafter\ifx\csname natexlab\endcsname\relax\def\natexlab#1{#1}\fi
\expandafter\ifx\csname bibnamefont\endcsname\relax
  \def\bibnamefont#1{#1}\fi
\expandafter\ifx\csname bibfnamefont\endcsname\relax
  \def\bibfnamefont#1{#1}\fi
\expandafter\ifx\csname citenamefont\endcsname\relax
  \def\citenamefont#1{#1}\fi
\expandafter\ifx\csname url\endcsname\relax
  \def\url#1{\texttt{#1}}\fi
\expandafter\ifx\csname urlprefix\endcsname\relax\def\urlprefix{URL }\fi
\providecommand{\bibinfo}[2]{#2}
\providecommand{\eprint}[2][]{\url{#2}}
\bibitem{OPERA11} The OPERA collaboration, arXiv:1109.4897 (2011).
\bibitem{adamson07} Adamson, P. et al. (MINOS Collaboration), 
Phys. Rev. D {\bf 76}, 072005 (2007).
\bibitem{amelino-camelia11} Amelino-Camelia, G. et al., arXiv:1109.5172 (2011).
\bibitem{tamburini11} Tamburini, F., Laveder, M. arXiv:1109.5445 (2011).
\bibitem{giudice11} Giudice, G. F., Sibiryakov, S., Strumia, A. arXiv:1109.5682 (2011).
\bibitem{dvali11} Dvali, G., Vikman, A. arXiv:1109.5685 (2011).
\bibitem{gubser11} Gubser, S. S. arXiv:1109.5687 (2011).
\bibitem{franklin11} Franklin, I. arXiv:1110.0234 (2011).
\bibitem{drago11} Drago, A., Masina, I., Pagliara, G., Tripiccione, R. arXiv:1109.5917 (2011).
\bibitem{magueijo11} Magueijo, J. arXiv:1109.6055 (2011).
\bibitem{cohen11} Cohen, A. G., Glashow, S. L. arXiv:1109.6562 (2011).
\bibitem{ciuffoli11} Ciuffoli, E., Evslin, J., Liu, J., Zhang, X. arXiv:1109.6641 (2011).
\bibitem{bi11} Bi, X.-J., Yin, P.-F., Yu, Z.-H., Yuan, Q. arXiv:1109.6667 (2011).
\bibitem{waxman97} Waxman, E., Bahcall, J. Phys. Rev. Lett. {\bf 78}, 2292 (1997).
\bibitem{gupta07} Gupta, N., Zhang, B. AstroParticle Phys. {\bf 27}, 386 (2007).
\bibitem{murase06} Murase, K. et al. Astrophys. J. {\bf 651}, L5 (2006).
\bibitem{meszaros01} M\'esz\'aros, P., Waxman, E.  Phys. Rev. Lett. {\bf 87}, 171102 (2001).
\bibitem{razzaque03} Razzaque, S, M\'esz\'aros, P., Waxman, E.  Phys. Rev. D {\bf 68}, 083001 (2003).
\end{thebibliography}
\end{document}